# Super Dielectric Materials

By Samuel Fromille and Jonathan Phillips*
Physics Department
Naval Postgraduate School
Monterey, CA 93943
*Email- jphillip@nps.edu

ABSTRACT

Evidence is provided that a class of materials with dielectric constants greater than $10^5$, herein called super dielectric materials (SDM), can be generated readily from common, inexpensive materials. Specifically it is demonstrated that high surface area alumina powders, loaded to the incipient wetness point with a solution of boric acid dissolved in water, have dielectric constants greater than $4*10^8$ in all cases, a remarkable increase over the best dielectric constants previously measured, ca. $1*10^4$. It is postulated that any porous, electrically insulating material (e.g. high surface area powders of silica, titania), filled with a liquid containing a high concentration of ionic species will potentially be an SDM. Capacitors created with the first generated SDM dielectrics (alumina with boric acid solution), herein called New Paradigm Super (NPS) capacitors display typical electrostatic capacitive behavior, such as increasing capacitance with decreasing thickness, and can be cycled, but are limited to a maximum effective operating voltage of about 0.8 V. A simple theory is presented: Water containing relative high concentrations of dissolved ions saturates all, or virtually all, the pores (average diameter 500 Å) of the alumina. In an applied field the positive ionic species migrate to the cathode end, and the negative ions to the anode end of each drop. This creates giant dipoles with high charge, hence leading to high dielectric constant behavior. At about 0.8 volts, water begins to break down, creating enough ionic species to 'short' the individual water droplets. Potentially NPS capacitor stacks can surpass 'supercapacitors' in volumetric energy density.

INTRODUCTION

There are several distinct capacitor technologies, and for understanding the import of the present work it is helpful to compare/contrast two types; supercapacitors, and traditional electrostatic (or 'ceramic') capacitors (1). In essence, supercapacitors increase capacitance by increasing electrically conductive electrode surface area. Most of the volume of a supercapacitor is the high surface area electrode in powder form. Increasing the supercapacitor volume while maintaining the shape/surface area of the attaching ends constant, increases the amount of electrode material area, concomitantly increasing the capacitance in direct proportion to the volume increase. Hence, supercapacitor performance is generally reported as 'per gram' or per unit volume. This explains the recent interest in employing graphene in supercapacitors (2-4), as graphene is arguably the 'ultimate' material for creation of high surface area electrodes. Indeed, it has very high electrical conductivity and the measured surface of some graphene forms are near the theoretical limit (~2700 $m^2$/g). Once the best supercapacitors incorporate graphene, supercapacitor energy density will be near a theoretical limit, hence further significant energy density increase for supercapacitors is unlikely.

In contrast to supercapacitors, the capacitance of a traditional electrostatic capacitor with constant sized electrodes decreases with volume. In an electrostatic capacitor, capacitance is inversely proportional to the distance between plates. Given plates of a constant size, the thinner an electrostatic capacitor, the greater the capacitance. Clearly, the means to improve the performance of this style of capacitor is either to make them thinner and/or to find materials with higher dielectric constants.

Relative to supercapacitor improvements, the ultimate energy density of traditional electrostatic capacitors has been modest over the last few decades. The biggest improvement resulted from new techniques that permit the fabrication of thinner ceramic dielectric layers. Enhancements in the dielectric constant of the best ceramic have been more modest. In fact, most of the effort to increase dielectric constant has focused on improving one material, barium titanate, for decades (5-9).

Herein, we introduce a novel hypothesis for a class of super dielectric materials (SDM), that is materials with dielectric constants greater than $10^5$, and provide test data demonstrating the existence of one SDM material. The hypothesis: *Charge species in liquid drops in the pores of solids will migrate to create dipoles, equal in size to the drops, in an applied electric field.* This phenomenology can be manipulated to create a high dielectric material, potentially to be deployed in a new generation of electrostatic capacitors, so called New Paradigm Super (NPS) capacitors. Specifically, we make the following 'application postulate': *Adding solutions containing ions (e.g. acid solutions) to highly porous insulating materials creates a high dielectric, or even 'superdielectric' (dielectric constant >$10^5$), material.*

Simple studies reported herein suggest this hypothesis is correct, and that the proposed application postulate works in practice. Specifically, the measured dielectric constant of one example, high surface area alumina incorporating a solution of boric acid, is orders of magnitude higher (ca. ~$10^9$) than any form of barium titanate. It is in fact an SDM. The high dielectric constants measured suggest a path forward to developing a classic electrostatic capacitor as an alternative to supercapacitors for electrical energy storage/power delivery.

The results also reveal a limitation to the initial design: The electrolyte 'breaks down', resulting in conduction, just as in a superconductor, at a relatively low voltage (10). For the particular materials combination reported here the ultimate 'capacitive' voltage was about 0.8 Volts. Still, extrapolating the current results to a 'depth' typical of ceramic capacitors (5 micron), and assuming better electrolytes will yield a higher ultimate voltage (2.5 V), yields an energy density for NPS capacitors of order 1000 J/cm$^3$, a value several times higher than that of the best commercial supercapacitor. In fact, these computations are conservative; for example commercial ceramic capacitors with dielectric thickness of <1μ are now available, even in large stacks (11).

EXPERIMENTAL

Dielectric Fabrication: The materials employed to create the specific dielectric employed in this study, alumina/boric acid solution super dielectric material (A-SDM), were high surface area aluminum oxide powder (Alfa Aesar, γ-phase, 99.97%, 3 micron APS Powder, S.A. 80-120 m$^2$/g, CAS 1344-28-1), boric acid powder (BDH, 99.5% $H_3BO_3$, CAS 10043-35-3), and distilled deionized water. These constituents were mixed by hand in this ratio in all cases: 1 g alumina: 1 mL H2O: 0.1g boric acid powder. This created a spreadable paste with no 'free' water (incipient wetness). It is interesting to note that a mixture of 1 mL of water and 0.1 g boric acid is only weakly acidic with a theoretical pH of approximately 4.5.

As pore structure is a significant component of the proposed model, the surface area and pore structure were determined from BET nitrogen isotherms collected at 77K and analyzed using a Quantachrome NOVA 4200e. Two samples were independently measured and both yielded results within 5% for all parameters; specifically a surface area of 39 +/-1 $m^2/g$, a total pore volume of 0.45 $cm^3/g$ and an average pore radius of 245 +/- 3 Å.

The dielectric paste was spread evenly on a 5 cm diameter disc of GTA grade Grafoil (0.76 mm thick, >99.99% carbon). As described elsewhere (12, 13) Grafoil is a commercially available high purity carbon material (available in sheets or rolls) made by compressing naturally occurring graphite flakes with a surface are on the order of 20 $m^2/g$. In the final step a second sheet of Grafoil is place on top, the thin 'capacitor' then mechanically pressed to create a near constant thickness as determined by measurements made at multiple positions using a hand held micrometer. The 'effective thickness' of the dielectric used in all computations herein was based on subtracting the Grafoil sheets thickness from the measured gross thickness of the capacitor.

Once constructed the capacitors were placed in an electrically insulating plastic jig with bottom and top cylindrical aluminum electrodes of 5 cm diameter and 5 mm thickness. A 250 g weight was placed on top in all cases. These capacitors were then placed in simple circuits (Figure 1) for measurements of charge and discharge. It is important to note that charging and discharging were generally done through several different resistors, specifically nominal 528 K, 99 K, or 20.1 kOhm resistors. The smallest was used in order to speed up processes, such as for multi-cycle tests, and for the thin capacitors in which the capacitance was greater than 50 milliFarads.

The primary test platform was a National Instruments ELVIS II electronics prototyping board implemented with LabView 2011 software. An additional multimeter, Agilent U1252A, was used for independent parameter verification. It is further notable that the capacitance of several types of commercial capacitors were measured using the above described instruments and protocol, and in every case the measured value and the listed value were within 30%.

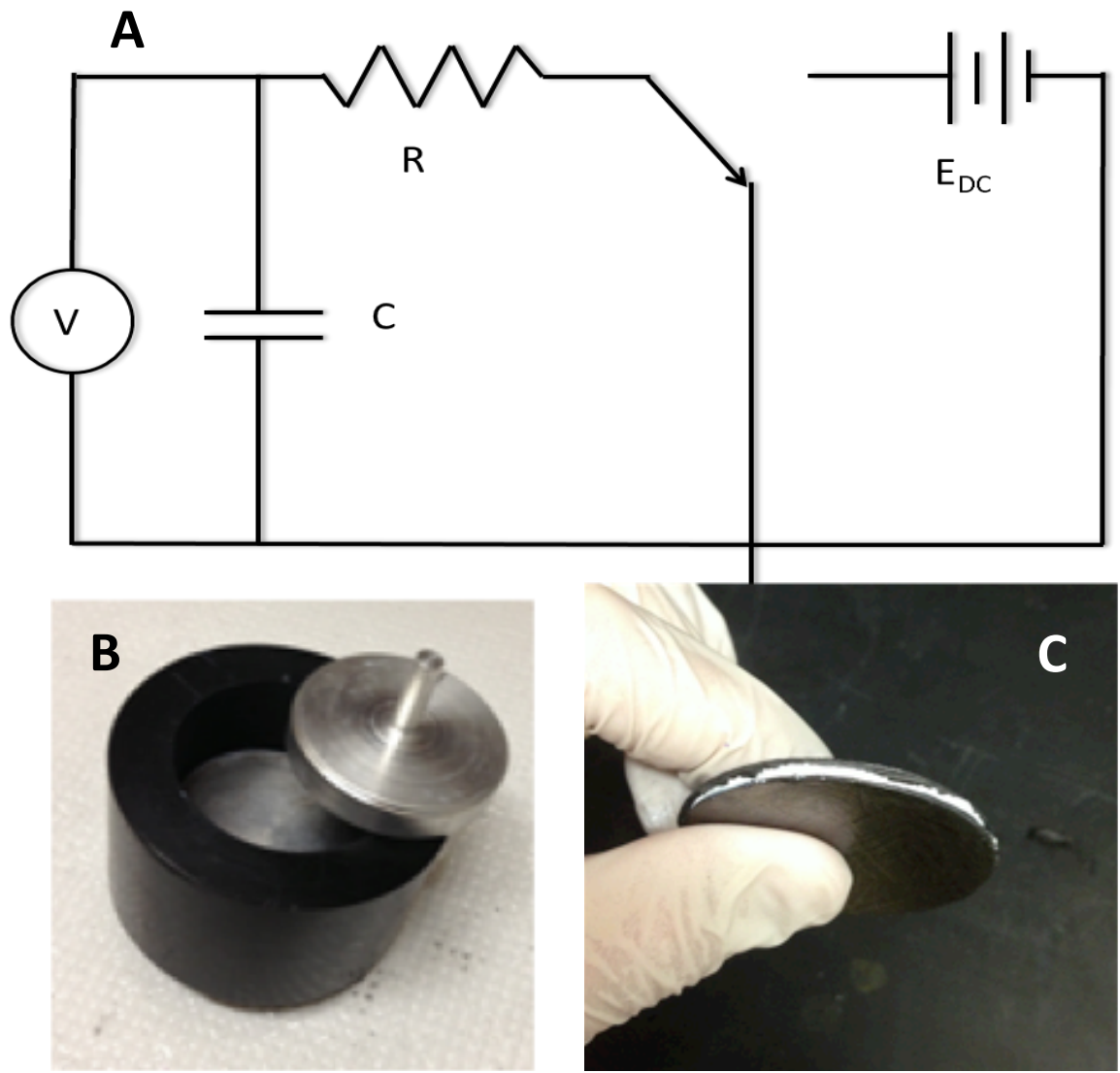

**Figure 1-** *Measuring Capacitive Properties*. A) With the switch 'down' the discharge voltage across the capacitor is measured, and with the switch 'up' the voltage across the capacitor during charging is recorded. B) The capacitor is placed in a hard plastic jig between two aluminum electrodes with diameter of 5 cm. C) A picture of the completed capacitor with the aluminum/water/boric acid paste squeezed between two sheets of Grafoil.

RESULTS

Basic phenomenological data from studies of cyclic charging and discharging show that capacitors employing A-SDM behave nearly as ideal capacitors over a limited, repeatable, voltage range. Typical multi-cycle data from one NPS capacitor, charged through a 99 kOhm resistor from a power supply operated at 4 V, then discharged through the same resistor, is shown in Figure 2. Re-plotting similar data from several NPS capacitors using A-SDM dielectric of different thicknesses, and discharged through a 528 KOhm resistor, was done to test the proposition that these capacitors charge/discharge exponentially, as per standard electrostatic capacitors:

$$\ln(V/V_0) = t/RC \qquad (1)$$

Moreover, the dielectric constant can be obtained from the time constant and this standard equation:

$$C = \varepsilon_0 \varepsilon_R \frac{A}{d} \qquad (2)$$

where $\varepsilon_0$ is the permittivity of free space (8.85 $10^{-12}$ F/m) and $\varepsilon_R$ is the dielectric constant. The area of the plate surface is $A$ and the distance between the two electrode surfaces is $d$.

From the plots of one multi-cycle data set (Figure 2) it is clear that in all cases below about 0.8 V the A-SDM have nearly constant time constants, hence constant capacitance. Using these measured time constants, the resistance value, and the physical parameters of the capacitors the dielectric constants were computed for all three charge and discharge cycles. Specifically, for the first discharge cycle the dielectric constant was ~$1.1*10^9$, but it roughly doubled by the third cycle. The first charge cycle showed a dielectric constant of ~$1.0*10^9$, but the last charge cycle dielectric constant was roughly only one third that value. Employing even the lower first cycle values shows that the A-SDM material is a super dielectric material. Its dielectric constants is orders of magnitude greater than that required by the definition of SDM: materials possessing a dielectric constant >$10^5$.

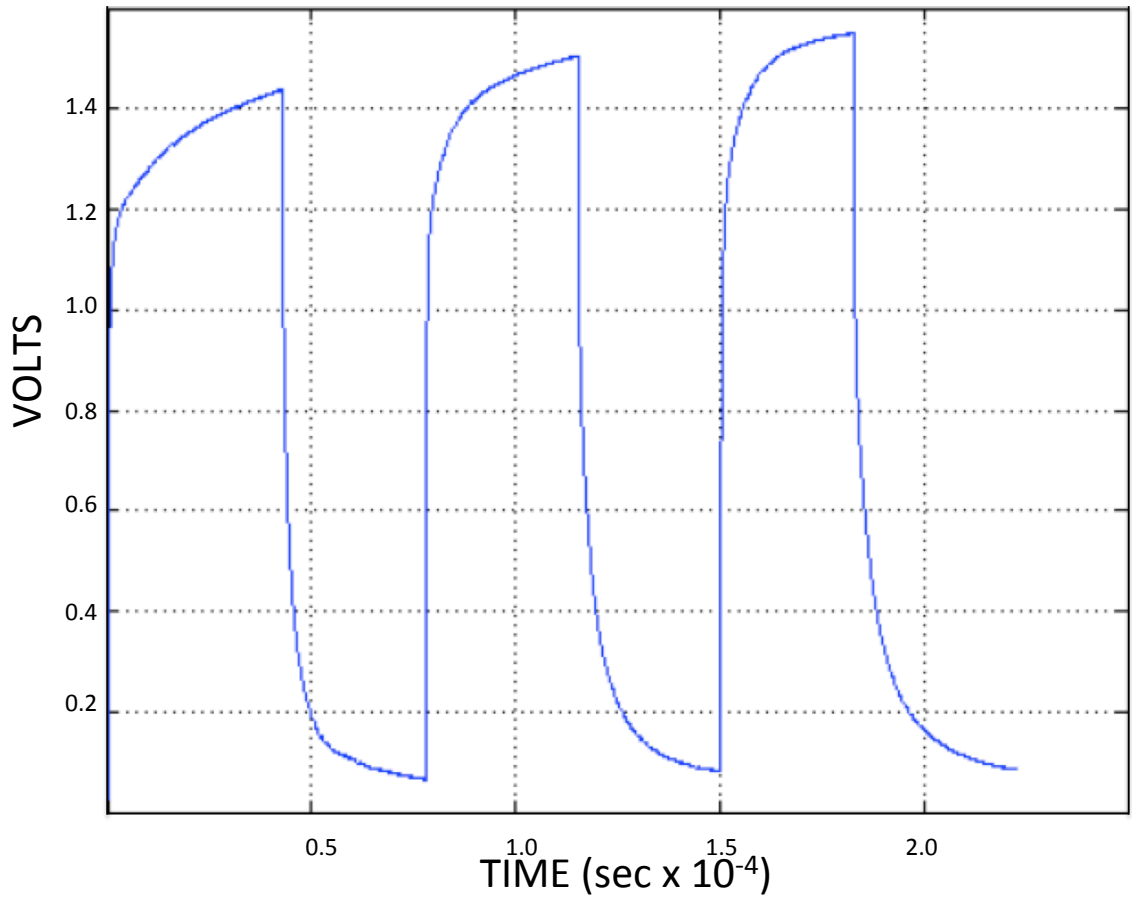

**Figure 2-** *Capacitive Cycling*. These capacitors go through regular cycles, as shown. Note that the discharge time during the steady 'capacitive part' of the cycles (below ~0.8 V) for a 99 kOhm resistor) is more than 2000 seconds. There is some difference between charge and discharge cycles in terms of apparent capacitance.

It is important to recognize limitations of 'Generation Zero' NPS-C created using A-SDM. First, they can only be charged to an ultimate voltage less than 2 Volts, no matter the applied charging voltage. Second, they only have a constant, high (SDM range) dielectric value, below about 0.8 V. (This is similar to the 'capacitive voltage' of a supercapacitor employing water as the electrolyte.) There is some small capacitance between the ultimate charging value (ca. 1.5 volts) and the onset of super dielectric behavior (0.8 V) but it does not contribute significantly to energy storage capacity. Third, these materials (Generation Zero) are moderately unstable, as clearly all major aspects of behavior including ultimate voltage and dielectric constant change moderately (ca. 10%) with each cycle. The magnitude of the instability is evident from an analysis of both the charging and discharging behavior of each part of the cycle. Using the highest voltage observed in each cycle for $V_0$, the charging and discharge behavior is plotted in Figure 3, as per Eq. 1. It is clear that the capacitive behavior changes with each cycle.

The observed 'instability', that is changing dielectric value, may result from gradual drying of the dielectric material. This postulate is consistent with one boundary condition repeatedly found: Approximately 10 days after creation, the capacitance of all samples went to zero. Physical examination of the dielectric material at this point showed it to be dry and cracked, in clear contrast to the initial wet, smooth, pasty consistency of the newly made capacitors. Moreover, it was repeatedly found that the careful addition of water to the dry, zero capacitance, dielectric material restored most of the original capacitive value. A simple hypothesis is consistent with the observations of i) an initially increasing dielectric constant, and ii) an eventual decrease to zero. To wit: There is an optimum amount, from the perspective of dielectric constant, of water in the A-SDM material.

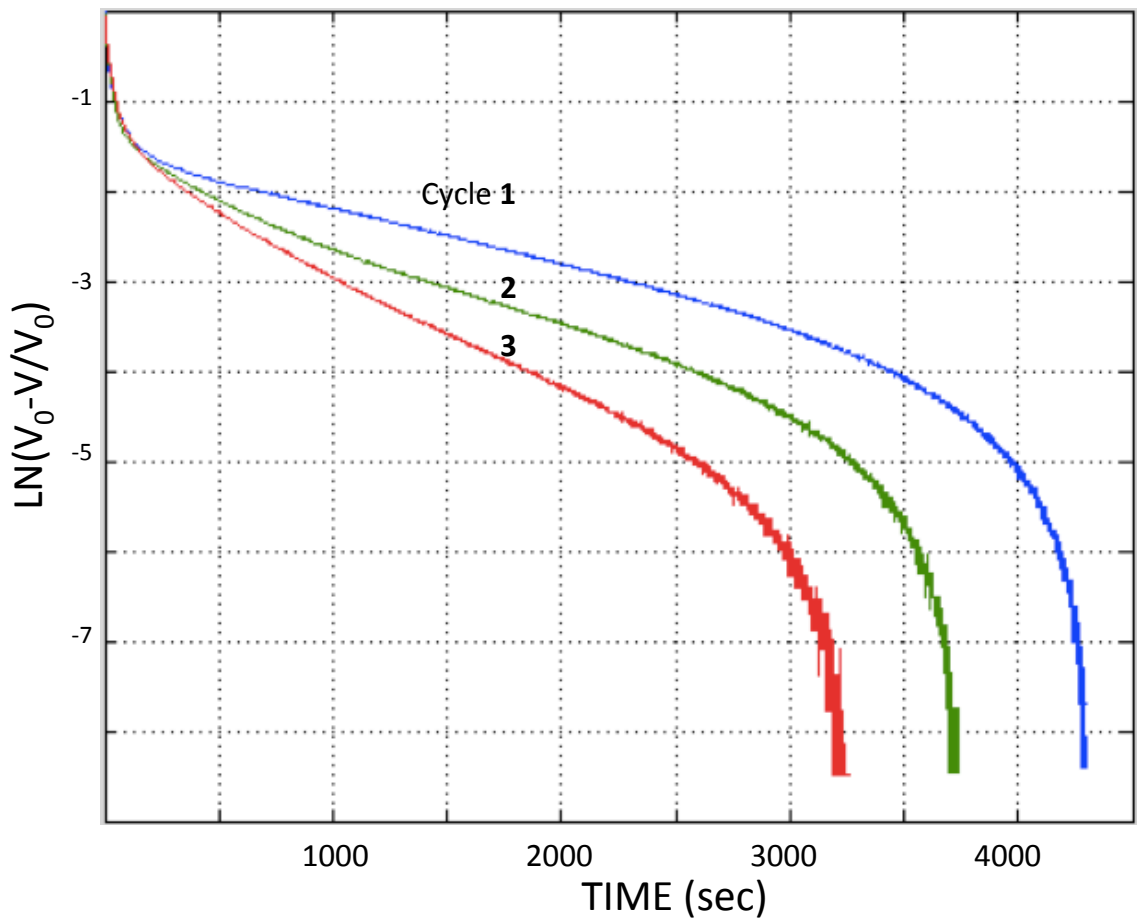

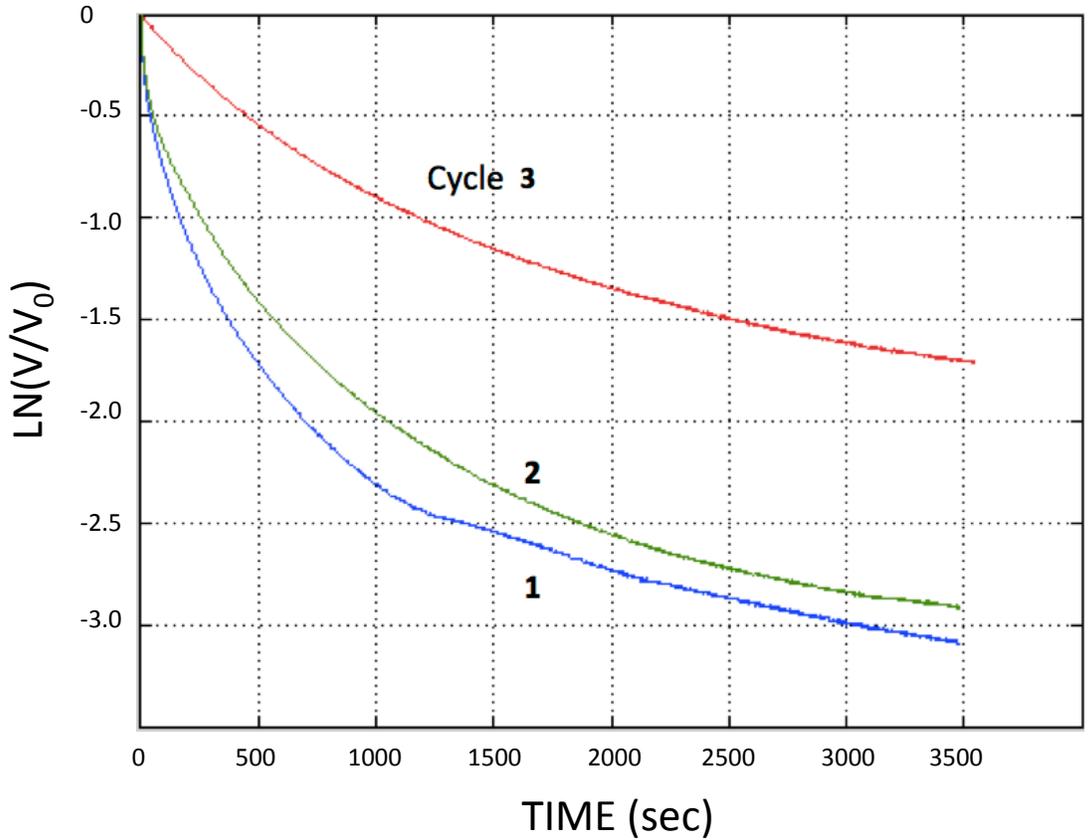

**Figure 3**- *Determination of Aging on Capacitive Behavior*. From Figure 2 cycle data charge and discharge constants can be derived. TOP- The charge cycle data from Figure 2 is replotted using Eq.1, and shows clear regions of constant capacitance during charging. BOTTOM- The same procedure was applied to the discharge data from Figure 2.

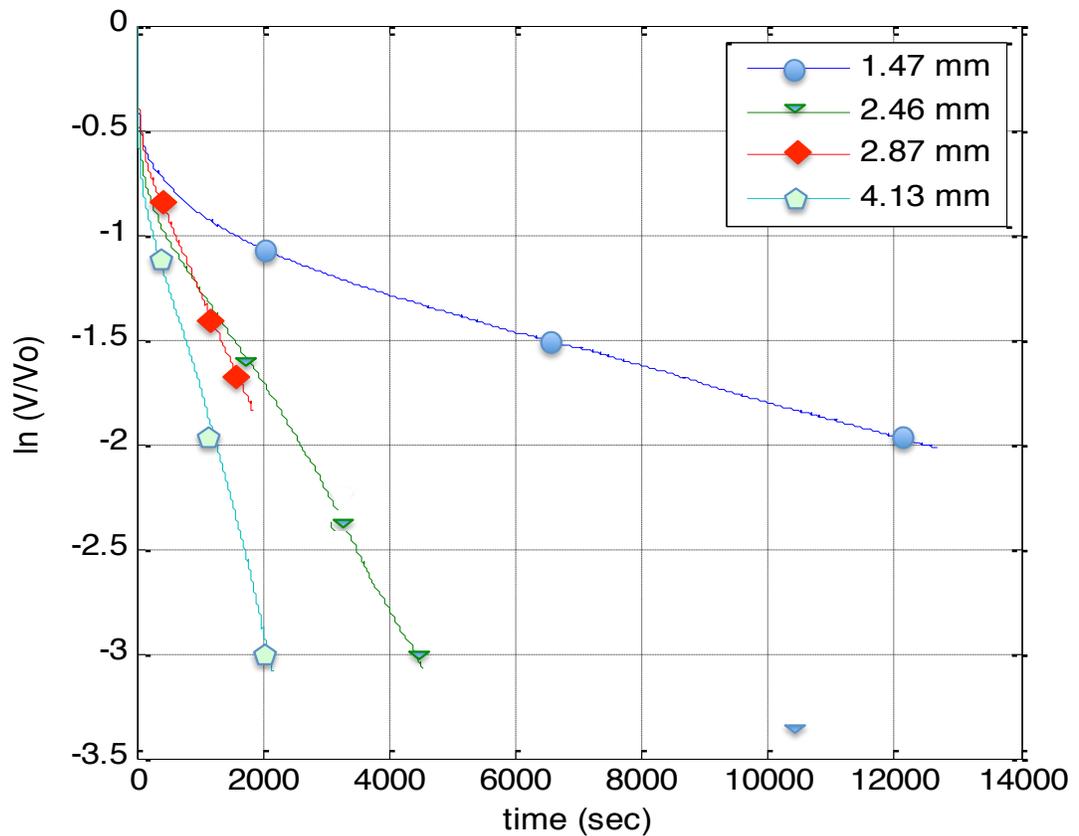

**Figure 4**- *Dielectric Constant through Large Load.* The discharge data through a 528 kOhm resistor for four alumina/boric acid SDM capacitors was fitted using Eq 2. It is clear that the linear part of a plot of $\ln(V/V_0)$ vs. time will yield the dielectric constant.

Discharge behavior for the thicker, hence lower total capacitance, samples was found to show behavior more like that anticipated for classic capacitors. That is, discharge through a 528 kOhm resistor, re-plotted per Eq. 1, was more linear (Figure 4) than that observed through a 99 kOhm resistor (Figure 3). This was not found to be the case for the thinner, higher capacitance samples. These samples displayed linear, constant dielectric behavior even for discharge through a 20 kOhm resistor.

Only one parameter of the NPS-C was systematically modified in this (initial) study: thickness of the A-SDM dielectric layer. As shown in Table I, NPS-C capacitors with seven different thicknesses of the dielectric material were created and tested. The plotted data (Figure 5) shows that the capacitance, first discharge cycle only, increases with decreasing thickness. From a fit of the line the capacitance can be projected to any thickness, and that value employed to predict energy density as a function of thickness.

| Test | Dielectric Thickness ($d$) | Initial Discharge Voltage ($V_0$) | Dielectric Constant ($\varepsilon_R$) at Operating Voltage | Operating Voltage | Dielectric Constant ($\varepsilon_R$) over Entire Range |
|---|---|---|---|---|---|
| Discharge Only (528kΩ) | 1.47 mm | 2.20 V | 1.81E9 | 0.7 V | 8.02E8 |
| Discharge Only (528kΩ) | 2.46 mm | 2.16 V | 5.78E8 | 0.8 V | 3.52E8 |
| Discharge Only (528kΩ) | 2.87 mm | 1.85 V | 4.44E8 | 0.9 V | 2.66E8 |
| Discharge Only (528kΩ) | 4.13 mm | 2.18 V | 4.43E8 | 0.8 V | 2.86E8 |
| Discharge Only (99kΩ) | 2.59mm | 1.43 V | 5.0E9 | 0.2 V | 1.2E9 |
| Discharge Only (20.8kΩ) | 0.38 mm | 1.60 V | 1.27E9 | 0.55 V | - |
| Discharge Only(20.8kΩ) | 0.25 mm | 1.44 V | 1.54E9 | 0.6 V | - |

**TABLE I-** *Impact of Thickness*. Data in the first four rows shown correspond to Figure 3, hence discharge was through a 528 KOhm resistor, and the last two rows through a 20 kOhm resistor. As shown the thinner the dielectric layer the higher the capacitance. It is also clear that the lower the 'resistor' the higher the capacitance. Variation in measured dielectric constant possibly reflects the irregularity of hand made construction.

From the fitted line we obtain a nominal dielectric constant for the particular alumina chosen, for the particular water/boric acid/alumina ratio employed in this work

(see experimental section), ~1*10⁹, clearly making this a 'superdielectric' below about 0.7 +/- 0.2 V.

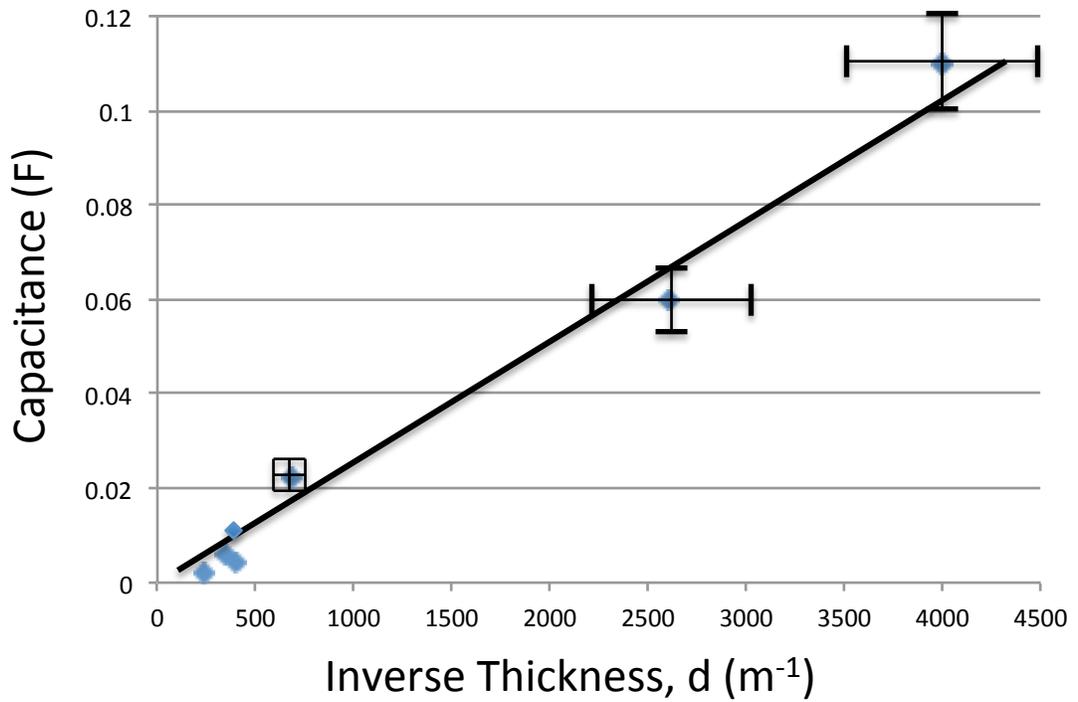

**Figure 5-** *Impact of Thickness.* Ideally (Equation 2), there is an inverse relationship between capacitance and thickness, d. As shown, a linear fit of the data is consistent with theory. Note: the error bars for the 'thick' dielectric capacitors are not shown as the errors are of the same size as the icons.

DISCUSSION

The most important result of this work is the demonstration of the existence of a new class of dielectrics, super dielectric materials, with dielectric constants greater than $10^5$. The specific dielectric employed was a high surface area alumina impregnated, to the consistency of paste, with an aqueous solution of weak boric acid (A-SDM). This material repeatedly showed dielectric constants greater than $4*10^8$. Thus, the data confirm the underlying application postulate: *Adding any ionic solution to highly porous insulating materials creates a high dielectric material*. Moreover, assuming a thickness of 5 micron, a typical value for inexpensive commercial ceramic capacitors, and using a rounded 'average' dielectric constant of $1*10^9$ (see Table I, Figure 5), yields a remarkable energy density of approximately 1,000 J/cm$^3$ at 2.5 volts. The voltage selected, 2.5 V, is typical for the breakdown of the best electrolytes. Even a computation based on the lowest dielectric constant ($4.4*10^8$) and the lowest 'capacitive voltage' (0.7 V) measured in the present work, leads to an energy density of ~40 J/cm$^3$, still a remarkable value.

The phenomenon observed are all consistent with the hypothesis/model stated in the Introduction: *Charge species in liquid drops in the pores of solids will migrate to create dipoles, equal in size to the drops, in an applied electric field*. The hypothesis was initially proposed on the basis of standard aspects of dielectric theory, in particular the understanding that the best solid dielectrics, such as barium titanate, are superior because of the magnitude and density of dipoles that form in an applied electric field. The greater the magnitude and density of dipoles, the better the dielectric. In the A-SDM it is postulated charge separation, leading to dipole formation, occurs in the nano-scale drops of ion containing liquid in the pores of the alumina. Specifically, in an applied electric field the positive ions will tend to cluster toward the capacitor cathode, and negative ions toward the anode. This creates a greater charge separation, physically longer, and probably larger, than possible in a solid crystal. Indeed, charge motion in a solid crystal is physically limited to a very small distance, in fact less than the diameter of an atom. In a water drop charge separation can occur over the entire length of a pore. Moreover, in solids the positive ions cannot move, whereas in a liquid both negatively and positively charged ions can migrate, increasing the dipole moment.

In particular, the model is consistent with these observations. First, at a voltage of around one volt, there is a 'dielectric breakdown' of water. As the water will not only exist in the pores, but will form a matrix that fills all empty space within the powder, once this breakdown occurs, there will be a conduction path from cathode to anode. Hence, at voltages above the breakdown voltage the effective dielectric constant will drop quickly to zero, as observed. Even absent an overall discharge path between electrodes, each individual drop would no longer be able to support a charge separation above the breakdown voltage. Second, in the absence of water there are no liquid drops available to form dipoles. This is consistent with the virtually disappearance of any capacitance once the dielectric fully dries. Third, once water is added to a desiccated dielectric the drops can reform, the chargeable species will still be present (probably on pore walls), and will re-dissolve, hence, the original dielectric behavior will be observed.

The above list of 'consistencies' are qualitative. A rough case can also be made that the large dielectric constants observed are quantitatively consistent with the model. In the case of the alumina employed herein, BET analysis indicated the average pore radius was order 250 Å. This indicates that the average dipoles are of length order 500 Å. As electric dipole moment is proportional to the charge separation distance, and potential energy is proportional to the magnitude of the dipole, energy is proportional to charge separation. Moreover, dipole moment is proportional to the amount of charge separated. The combination of three orders of magnitude increase in dipole length and a significant increase in the magnitude of the charge separation could explain the 4 or 5 order of magnitude increase in dielectric constant observed relative to that found for barium titanate.

Clearly the quantitative agreement is 'approximate'. It is not possible with current data to provide precise values of the charge concentrations in the liquid, nor is it possible to determine other relevant parameters such as actual concentrations of charge at either end of a drop. However, some speculative comments are useful. In particular, it is possible that on this scale each drop acts as a nearly perfect conductor, such that charges move until the field inside the drop is cancelled. This would make the drops 'metal like', with nearly infinite permittivity. The combination of very large physical dipoles, of nearly infinite permittivity, would create the super dielectric constant values observed in the present work. This postulate will be tested in future work via the measurement of capacitance as a function of temperature. If correct, the capacitance should decrease with increasing temperature, and increase with decreasing temperature, until the drops solidify (freeze), at which point the dielectric constant should drop dramatically.

Other aspects of the model can be tested as well. For example the model suggests larger pores may produce higher dielectric values. Certainly, there are many alumina materials with different pore sizes readily available for testing this objectively. It is possible that the addition of mercury to a material with large pores will create a super dielectric material as mercury, being a metal, has nearly an infinite permittivity.

Many objections to the proposed model will inevitably be raised. One likely objection is that water cannot enter pores below a certain size due to surface tension. Technically, this is correct, however; it is well known that water in the form of water vapor will enter pores of any size. In the event one or more 'primary adsorption sites' (PAH) exist in the pore, the water molecule will adsorb and nucleate the formation of a drop of water from other vapor phase water molecules. This leads to Type III isotherm behavior and the complete filling of the pore at a vapor pressure equal about 50% relative humidity (14-16). As the wetted alumina should have a local RH of nearly 100%, and there is a high density of PAH on a hydrophilic alumina surface, all pores, any size, should be filled. Future studies may test the SDM hypothesis through the use of alumina treated to create a hydrophobic surface. Such material should not exhibit SDM behavior.

Finally, it is interesting to speculate on the potential value of NPS capacitors. As noted earlier, with reasonable extrapolation of the collected data, and the projected use of a dielectric with a higher discharge voltage, leads to a remarkable energy density of ~1000 J/cm3. A D-battery ('flashlight battery') has a volume of ~53 cm$^3$. Assuming that about half that volume is taken up by SDM of 5 micron thickness, the rest of the volume evenly divided between electrode and insulating layers, means a l D-cell sized NPS capacitor could hold 25,000 J. In contrast, the best 'd-cell' supercapacitors (costly) advertise a capacitance of 3000 F and a voltage of 2.7 V, for a total energy of approximately 11,000 J. Could an NPS capacitor compete against a battery? A typical D battery can deliver just over 80,000 J. This in turn suggests that with optimization, NPS capacitors, made of remarkably inexpensive material, could surpass supercapacitors, and rival batteries in terms of volumetric energy density.